\documentclass[11pt]{article}
\usepackage{epsfig}
\usepackage{amssymb}
\usepackage{amsthm}
\usepackage{amscd}
\usepackage{amsfonts}
\usepackage{amsmath}
\usepackage[T1]{fontenc}
\usepackage{ae,aecompl}
\usepackage{pslatex}
\usepackage{graphicx}
\usepackage{color}
\usepackage{latexsym}  
\usepackage{url}
\usepackage{setspace}

\newtheorem{thm}{Theorem}

\newtheorem{defn}[thm]{Definition}

\newtheorem{prop}[thm]{Proposition}

\begin{document}
\title{Parametric Inference for Biological Sequence Analysis}

\author{Lior Pachter and Bernd Sturmfels \\
Department of Mathematics, University of California, Berkeley, CA
94720}

\maketitle

\begin{abstract}
One of the major successes in computational biology has been the 
unification, using the graphical model formalism, of a multitude of 
algorithms for annotating and comparing biological sequences.
Graphical models that have been applied towards these problems include 
hidden Markov models for annotation, tree 
models for phylogenetics, and pair hidden Markov models for 
alignment. A single algorithm, the sum-product algorithm, solves
many of the inference problems associated with different statistical models. 
This paper introduces the \emph{polytope propagation algorithm}
for computing the Newton polytope of an observation from a graphical model.
This algorithm is a geometric version of the sum-product algorithm and 
is used to analyze the parametric behavior of maximum a posteriori 
inference calculations for graphical models.
\end{abstract}


\section{Inference with Graphical Models for Biological Sequence Analysis}

This paper develops a new
 algorithm for graphical models based on the mathematical foundation for statistical models proposed in \cite{Pachter:04}. Its relevance 
for  computational biology can be summarized as follows:

\textbf {(a) Graphical models are a unifying statistical framework for biological sequence analysis.}

\textbf {(b) Parametric inference is important for obtaining biologically meaningful results.}

\textbf {(c) The polytope propagation algorithm solves the parametric inference problem.}

\vskip .1cm

Thesis (a) states that graphical models are good models for biological sequences. This emerging understanding is the result of practical success with 
probabilistic algorithms, and also the observation that inference algorithms for graphical models subsume many apparently non-statistical methods.
 A noteworthy example of the latter is the explanation of classic alignment
algorithms such as Needleman-Wunsch and Smith-Waterman in terms of the Viterbi algorithm for pair hidden Markov models \cite{Bucher:96}.
Graphical models are now used for many problems including motif detection, gene finding, alignment, phylogeny reconstruction and protein structure prediction. For example, most gene prediction methods are now hidden Markov model (HMM) based, and previously non-probabilistic methods 
now have HMM based re-implementations.

In typical applications, biological sequences are modeled as {\em observed random variables} $Y_1,\ldots,Y_n$ in a graphical model. The observed random variables may correspond to sequence elements such as nucleotides or amino acids. {\em Hidden random variables} $X_1,\ldots,X_m$ encode information of interest that is unknown, but which one would like to infer. For example, the information could be an annotation, alignment or ancestral sequence in a phylogenetic tree. One of the strengths of graphical models is that by virtue of being probabilistic, they can be combined into powerful models where the hidden variables are more complex. For example, hidden Markov models can be combined with pair hidden Markov models to simultaneously align and annotate sequences \cite{Alexandersson:03}. One of the drawbacks of such approaches is that the models have more parameters and as a result inferences could be less robust.

For a fixed observed sequence $\sigma_1 \sigma_2 \ldots \sigma_n$ and {\em fixed parameters}, 
the standard inference problems are:
\begin{enumerate}
\item[1.] the calculation of {\em marginal probabilities}:
\[ p_{\sigma_1 \cdots \sigma_n}
\quad = \quad 
\sum_{h_1,\ldots,h_m} {\rm Prob} (X_1=h_1,\ldots,X_m=h_m,Y_1=\sigma_1,\ldots,Y_n=\sigma_n) \]
\item[2.] the calculation of {\em maximum a posteriori log probabilities}:
\[ \delta_{\sigma_1 \cdots \sigma_n}
\quad = \quad
 \min_{h_1,\ldots,h_m} - {\rm log} \left( {\rm Prob} (X_1=h_1,\ldots,X_m=h_m,Y_1=\sigma_1,\ldots,Y_n=\sigma_n) \right), \]
\end{enumerate}
where the $h_i$ range over all the possible assignments for the hidden random variables $X_i$. 
In practice, it is the solution to Problem 2 that is of interest, since it is the one that solves the problem of finding the genes in a genome or the ``best'' alignment for a pair of sequences. 
A shortcoming of this approach is that the solution $\widehat {\bf h} = (\hat h_1, \ldots, \hat h_m)$ may vary considerably with a change in parameters. 

Thesis (b) suggests that a {\em parametric} solution to the inference problem can help in ascertaining the reliability, robustness and biological meaning of an inference result. By {\em parametric inference} we mean the solution of 
Problem 2 for all model parameters simultaneously. In this way we can decide if a solution
obtained for particular parameters is an artifact or is largely independent of the chosen 
parameters. This approach has already been applied successfully to the problem of pairwise sequence alignment in which parameter choices are known to be crucial in obtaining good alignments \cite{Fernandez-Baca:00, Gusfield:96, Waterman:92}.
Our aim is to develop this approach for arbitrary graphical models.
In thesis (c) we claim that the polytope propagation algorithm is efficient for solving the parametric inference problem, and, in certain cases is not much slower than solving Problem 2 for fixed parameters.
The algorithm is a geometric 
version of the sum-product algorithm, which is the standard tool for
inference with graphical models.

The mathematical setting for understanding
the polytope propagation algorithm is {\em tropical geometry}.
The connection between tropical geometry and parametric inference  in statistical models
is developed in the companion paper \cite{Pachter:04}. Here we describe the details of the polytope propagation algorithm (Section 3) in two familiar settings: the hidden Markov model for annotation (Section 2) and the pair hidden Markov model for alignment (Section 4). Finally, in Section 5, we discuss some practical aspects of parametric inference, such as specializing parameters, the construction of single cones which eliminates the need for identifying all possible maximum a posteriori explanations, and the relevance of our findings to Bayesian computations.

\section{Parametric Inference with Hidden Markov Models}
Hidden Markov models play a central role in sequence analysis, 
where they are widely used to annotate DNA sequences \cite{Baldi:98}.
A simple example  is the CpG island annotation problem \cite[\S 3]{Durbin:98}.
CpG sites are locations in DNA sequences where 
the nucleotide cytosine (C) is situated next to a guanine (G) nucleotide (the ``p'' comes from the fact that a phosphate links them together). There are regions with many CpG sites in eukaryotic 
genomes, and these are of interest because of the action of DNA methyltransferase, which  
recognizes CpG sites and converts the cytosine  into 5-methylcytosine. Spontaneous deamination 
causes the 5-methylcytosine to be converted into thymine (T), and the mutation is not fixed 
by DNA repair mechanisms. This results in a gradual erosion
of CpG sites in the genome. {\em CpG islands} are regions of DNA with many unmethylated CpG sites. Spontaneous deamination of cytosine to thymine in these sites is repaired, resulting
in a restored CpG site. The computational identification of CpG islands is important, because they are associated with promoter regions of genes, and are known to be involved
in gene silencing.  

Unfortunately, there is no sequence characterization of CpG islands. A generally accepted definition due to Gardiner-Garden and Frommer \cite{Gardiner-Garden:87} 
is that a CpG island is a region of DNA at least 200bp long with a G+C content of at least 50\%, and with a ratio of observed to expected CpG sites of at least 0.6. This arbitrary
definition has since been refined (e.g. \cite{Takai:02}), however even analysis of the complete sequence of the human genome \cite{Lander:01} has failed to 
reveal precise criteria for what constitutes a CpG island. Hidden Markov models can be used to predict CpG islands \cite[\S 3]{Durbin:98}. We have selected this application of HMMs 
in order to illustrate our approach to parametric inference in a mathematically simple setting.

The CpG island HMM we consider has $n$ hidden binary random variables $X_i$, and $n$ observed random variables $Y_i$ that take 
on the values $\{A,C,G,T\}$ (see Figure 1 in \cite{Pachter:04}). In general, an
HMM can be characterized by the following conditional 
independence statements for  $i = 1 , \ldots,n$:
\begin{eqnarray*} & p(X_i \, | \,X_1,X_2,\ldots,X_{i-1}) \quad 
= \quad  p(X_i \,| \, X_{i-1}), 
\\& p(Y_i \, |\, X_1,\ldots,X_i,Y_1,\ldots,Y_{i-1})\quad = 
\quad p(Y_i \,|\, X_i). \end{eqnarray*}
The CpG island HMM has twelve model parameters, namely, the
entries of the transition matrices
$$ S \, = \, \begin{pmatrix}
s_{00} & s_{01} \\
s_{10} & s_{11} \\
\end{pmatrix}
\qquad \hbox{and} \qquad
T \, = \, \begin{pmatrix}
t_{0A} & t_{0C} & t_{0G} & t_{0T} \\
t_{1A} & t_{1C} & t_{1G} & t_{1T} 
\end{pmatrix}.
$$
Here the hidden state space has just two states non-CpG $=0$ and CpG $=1$ 
with transitions allowed between them, but in more complicated applications, such as gene finding, 
the state space is used to model numerous gene components (such as introns and exons) and 
the sparsity pattern of the matrix $S$ is crucial. In its algebraic representation 
\cite[\S 2]{Pachter:04}, the HMM is given as the image
of the polynomial map
\begin{equation}
\label{polymap}
f \, : \, {\bf R}^{12} \rightarrow {\bf R}^{4^n}, \,\,\,
(S,T) \ \mapsto \ \sum_{h \in \{0,1\}^n}  \ \  t_{h_1 \sigma_1}
s_{h_1 h_2} t_{h_2 \sigma_2} s_{h_2 h_3} \cdots
s_{h_{n-1} h_n} t_{h_n \sigma_n}.
\end{equation}
The inference problem 1 asks for an evaluation of one coordinate polynomial $f_\sigma$ of the map $f$. This can be done in linear time (in $n$) using the 
\emph{forward algorithm} \cite{Jordan:02}, 
which  recursively evaluates the formula
\begin{equation}
\label{sum-product}
 f_{\sigma} \quad = \quad
\sum_{h_n=0}^1 t_{h_n \sigma_n} \biggl(
\sum_{h_{n-1}=0}^1 s_{h_{n-1} h_n} t_{h_{n-1} \sigma_{n-1}}
\cdots
\bigl(
\sum_{h_2=0}^1 t_{h_2 h_3} s_{h_2 \sigma_2} 
(\sum_{h_1=0}^1 t_{h_1 h_2} s_{h_1 \sigma_1} )\bigr) \cdots \biggr)
\end{equation}
Problem 2 is to identify the largest term in the expansion of $f_\sigma$.
Equivalently, if we write $u_{ij} = - {\rm log}(s_{ij})$ and
$v_{ij} = - {\rm log}(t_{ij})$ then Problem 2 is to evaluate the piecewise-linear function
\begin{equation}
\label{ref:Viterbi}
 g_{\sigma} \,\, = \,\,
{\rm min}_{h_n} v_{h_n \sigma_n} + \bigl( 
{\rm min}_{h_{n-1}} u_{h_{n-1} h_n} + v_{h_{n-1} \sigma_{n-1}} + 
\cdots + 
\bigl(
{\rm min}_{h_2} v_{h_2 h_3} + u_{h_2 \sigma_2} +
( {\rm min}_{h_1} u_{h_1 h_2} + v_{h_1 \sigma_1} )\bigr) \cdots \ \bigr).
\end{equation}
This formula can  be efficiently evaluated by recursively computing the
parenthesized expressions. This is known as the
\emph{Viterbi algorithm} in the HMM literature. 
The Viterbi and forward algorithms are instances of 
the more general {\em sum-product algorithm} \cite{Kschischang:01}.

What we are proposing in this paper is to compute 
the collection of cones in ${\bf R}^{12}$
on which the piecewise-linear function $g_\sigma$ is linear. 
This may be feasible because the number of cones grows polynomially in $n$.
Each cone is indexed by
a  binary sequence ${\bf h} \in \{0,1\}^n$ which represents the CpG islands found
for any system of parameters $(u_{ij}, v_{ij})$ in that cone. A binary sequence which
arises in this manner is an \emph{explanation for $\sigma$} in the sense of 
\cite[\S 4]{Pachter:04}.
Our results in \cite{Pachter:04} imply that the number of explanations
scales polynomially with $n$.

\begin{thm}
For any given DNA sequence $\sigma$ of length $n$, the 
number of bit strings $\widehat {\bf h} \in \{0,1\}^n$ which are
explanations for the sequence $\sigma$ in the CpG island HMM
is bounded above by a constant times $n^{5.25}$.
\end{thm}

\begin{proof}
There are a total of $2 \cdot 4 + 4 = 12$ parameters which is the dimension of the
ambient space. Note, however, that for a fixed observed sequence the number of times 
the observation $A$ is made is fixed, and similarly for $C,G,T$. Furthermore, the total
number of transitions in the hidden states must equal $n$. Together, these constraints remove
five degrees of freedom. We can thus apply \cite[Theorem 7]{Pachter:04}
with $d=12-5 = 7$. This shows that
the total number of vertices of the Newton polytope of $\,f_{\bf \sigma}\,$ is
 $\,O(n^{\frac{7 \cdot 6}{8}}) = O(n^{5.25})$.
\end{proof}

\begin{figure}[ht]  
\begin{center}  
 \includegraphics[scale=0.35]{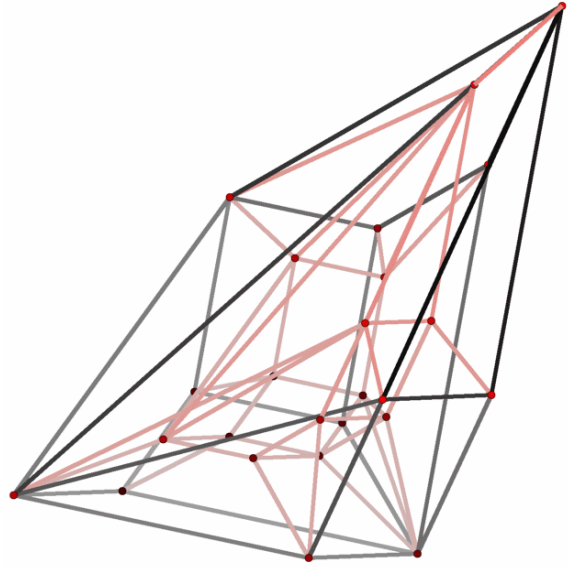} 
 \end{center}  
\caption{The Schlegel diagram of the Newton polytope of
 an observation in the CpG island HMM.}
\label{fig:Newton_polytope}
\end{figure}

 We explain the biological meaning of our parametric analysis
 with a very small example. 
 Let us consider the
 following special case of the CpG island HMM.
First, assume that $t_{iA}=t_{iT}$ and that $t_{iC}=t_{iG}$, i.e.,
the output probability depends only on whether the nucleotide
is a purine or pyrimidine. Furthermore, assume that
$t_{0A}=t_{0G}$, which means that the probability of emitting
a purine or a pyrimidine in the non-CpG island state is equal 
(i.e. base composition is uniform in non-CpG islands). 

 Suppose that the observed sequence is
${\bf \sigma}=AATAGCGG$. We ask for  {\em all} 
the possible explanations for ${\bf \sigma}$,
that is, for all possible maximum a posteriori 
CpG island annotations for all parameters.
A priori, the number of explanations is bounded by $2^8 = 256$, the total
number of binary strings of length eight. However, of the
$256$ binary strings, only $25$ are explanations.
Figure 1 is a geometric representation of the 
solution to this problem: the Newton polytope of $f_\sigma$ is
 a $4$-dimensional polytope with $25$ vertices.
The figure is a \emph{Schlegel diagram} of this polytope.
It was drawn with the software POLYMAKE 
 \cite{Gawrilow:00,Gawrilow:01}.
The $25$ vertices in Figure 1 correspond to the
$25$ annotations, which are the explanations for $\sigma$
as the parameters vary. Two annotations are connected by 
an edge if and only if their parameter cones share a wall. 
From this geometric representation, we can determine all
parameters which result in the 
same maximum a posteriori prediction.

\section{Polytope Propagation}

The evaluation of $g_{\sigma}$ for fixed parameters using the formulation in (\ref{ref:Viterbi}) is known as the Viterbi algorithm in the HMM literature. We begin by re-interpreting this algorithm as a convex optimization problem. 

\begin{defn}
The Newton polytope of a polynomial 
\[ f(x_1,\ldots,x_d) \quad = \quad \sum_{i=1}^{n} c_i \cdot x_1^{a_{1,i}} x_2^{a_{2,i}} \cdots x_d^{a_{d,i}} \]
is defined to be the convex hull of the lattice points in ${\bf R}^d$ corresponding to 
the monomials in $f$: 
\[ NP(f) \quad  = \quad
 conv\{(a_{1,1},a_{2,1},\ldots,a_{d,1}), \cdots, (a_{1,n},a_{2,n},\ldots,a_{d,n})\}. \]
\end{defn}
Recall that for a fixed observation there are natural polynomials associated with a graphical model, which we have been denoting by $f_{\sigma}$.
In the CpG island example from Section 2, these polynomials are the coordinates 
 $f_\sigma$ of the polynomial map $f$ in (\ref{polymap}).
 Each coordinate polynomial $f_\sigma$ is the sum of $2^n$ monomials,
 where $n = |\sigma|$. The crucial observation is that even though the number of monomials grows exponentially with $n$, the number of vertices of the
 Newton polytope $NP(f_\sigma)$ is much smaller. The Newton polytope
 is important for us because its vertices represent the solutions to the
 inference problem 2.

 \begin{prop}
\label{polytopepropagation}
The maximum a posteriori log probabilities $\,\delta_{\sigma}\,$ 
 in Problem 2 can be determined by 
minimizing a linear functional over the Newton polytope of $\,f_\sigma$.
\end{prop}

 \begin{proof}
This is nothing but a restatement of the fact that when passing to logarithms, monomials in the parameters become linear functions in the logarithms of the parameters. \end{proof}

Our main result in this section is an algorithm which we state
in the form of a theorem.

\begin{thm}[Polytope propagation]
Let $f_{\sigma}$ be the polynomial associated to a fixed observation $\sigma$ from a graphical model. The list of all vertices of the Newton polytope of $f_{\sigma}$ can be 
computed efficiently by recursive convex hull and Minkowski sum computations on unions of polytopes. 
\end{thm}

\begin{proof}
Observe that if $f_1,f_2$ are polynomials then $NP(f_1 \cdot f_2) = NP(f_1) + NP(f_2)$
where the $+$ on the right hand side denotes the Minkowski sum of the two
polytopes. Similarly, $\,NP(f_1+f_2) = {\rm conv} \bigl( NP(f_1) \cup NP(f_2) \bigr)\,$
if $f_1$ and $f_2$ are polynomials with positive coefficients.
The recursive description of $f_{\bf \sigma}$ given in (\ref{sum-product}) can be used
to evaluate the Newton polytope efficiently. The necessary geometric 
primitives are precisely Minkowski sum and convex hull of unions of convex polytopes. 
These primitives run in polynomial
time since the dimension of the polytopes is fixed. This is the
case in our situation since we consider graphical models
with a fixed number of parameters. We can hence
run the sum-product algorithm efficiently in the 
semiring known as the \emph{polytope algebra}.
The size of the output scales polynomially by \cite[Thm.~7]{Pachter:04}.
\end{proof}

\begin{figure}[ht]  
\begin{center}  
 \includegraphics[scale=0.75]{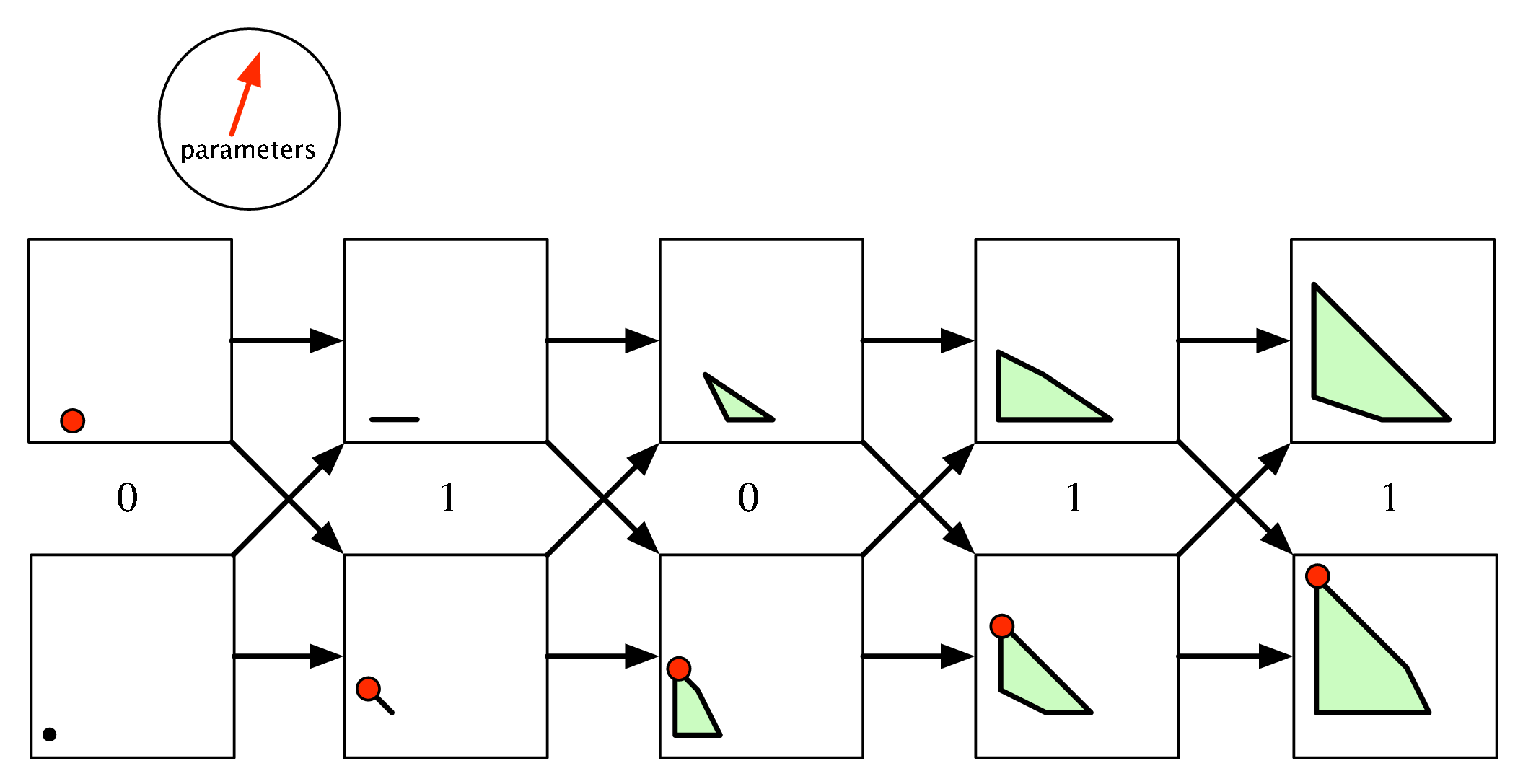} 
 \end{center}  
\caption{Graphical representation of the polytope propagation algorithm for a hidden Markov model.
For a particular pair of parameters, there is one
optimal Viterbi path (shown as large vertices on the polytopes).}
    \label{fig:HMMpoly}
\end{figure}

Figure 2 shows an example of the polytope propagation algorithm for a hidden Markov model
with all random variables binary and with the following transition and output
matrices:
$$ S \, = \, \begin{pmatrix}
s_{00} & 1 \\
1 & s_{11} \\
\end{pmatrix}
\qquad \hbox{and} \qquad
T \, = \, \begin{pmatrix}
s_{00} & 1  \\
 1 & s_{11} 
\end{pmatrix}.
$$
Here we specialized to only two parameters in order to simplify the diagram.
When we run polytope propagation for long enough DNA sequences
$\sigma$ in the 
CpG island HMM of Section 2 with all $12$ free parameters, we get a diagram just like Figure 2,
but with each polygon replaced by a seven-dimensional polytope.

It is useful to note that for HMMs, the Minkowski sum operations are simply shifts of the polytopes, and therefore the only non-trivial geometric operations required are the convex hulls of unions of polytopes.
The polytope in Figure 1 was computed using polytope propagation. This polytope 
has dimension $4$ (rather than $7$) because the sequence ${\bf \sigma}=AATAGCGG$ is so short.
We wish to emphasize that the small size of our examples is only for clarity; there is no practical
or theoretical barrier to computing much larger instances.

For general graphical models, the running time of the Minkowski sum and convex hull computations depends on the number of parameters, and the number of vertices in each computation. These are
clearly bounded by the total number of vertices of $NP(f_{\sigma})$, which are bounded above by \cite[Theorem 7]{Pachter:04}:
$$ \# \,{\rm vertices} (NP(f_\sigma)) \,\,\, \leq \,\,\,
  {\rm constant} \cdot E^{d(d-1)/(d+1)} \,\,\, \leq \,\,\, {\rm 
constant} \cdot E^{d-1} . $$
Here $E$ is the number of edges in the graphical model (often linear in the number of vertices of the model). The dimension $d$ of the Newton polytope $NP(f_\sigma)$
is fixed because it is bounded above by the number of model parameters.
The total running time
of the polytope propagation algorithm can then be estimated by multiplying the running time for the geometric operations of Minkowski sum and convex hull  
with the running time of the sum-product algorithm. In any case,
 the running time scales polynomially in $E$.

We have shown in \cite[\S 4]{Pachter:04}
that the vertices of $NP(f_{\sigma})$ correspond to explanations
for the observation $\sigma$. In parametric inference we are interested
in identifying the parameter regions that lead to the same explanations. 
Since parameters can be identified 
with linear functionals, it is the case that the set of parameters that lead to the same explanation (i.e. a vertex $v$) are those linear functionals that minimize on $v$. The
set of these linear functionals is the {\em normal cone of
$NP(f_\sigma)$ at $v$}. The collection of all normal cones 
at the various vertices $v$ forms the {\em normal fan} of the polytope. Putting this together with Proposition \ref{polytopepropagation} we obtain:

\begin{prop}
The normal fan of the Newton polytope of $f_{\sigma}$ solves the parametric
inference problem for an observation $\sigma$ in a graphical model.
It is computed using the polytope propagation algorithm.
\end{prop}

An implementation of polytope propagation for arbitrary graphical models
is currently being developed within the
geometry software package POLYMAKE \cite{Gawrilow:00,Gawrilow:01} by Michael Joswig.

\section{Parametric Sequence Alignment} 

The \emph{sequence alignment} problem asks to find the best alignment between two sequences which have evolved from a common ancestor via a series of mutations, insertions and deletions. Formally,
 given two sequences $\,\sigma^1 = 
 \sigma^1_1 \sigma^1_2 \cdots \sigma^1_n \,$ and
$\,\sigma^2 =   \sigma^2_1 \sigma^2_2 \cdots \sigma^2_m \,$ 
over the alphabet $ \{0,1,\ldots,l-1\}$,
 an \emph{alignment} is a string over the alphabet $\{M,I,D\}$ such that 
$\#M+\#D= n$ and $\#M+\#I=m $.
Here $\#M, \#I, \#D$ denote the number of characters $M,I,D$ 
in the word respectively.  An alignment records the ``edit steps'' from the sequence
$\sigma^1$ to the sequence $\sigma^2$, where edit operations consist of changing characters, 
preserving them, or inserting/deleting them. An $I$ in the alignment string 
corresponds to an insertion in the first sequence, a $D$ is a deletion in the first 
sequence, and an $M$ is either a character change, or lack thereof.
We write ${\cal A}_{n.m}$ for the set of all alignments.
For a given $h \in {\cal A}_{m,n}$, we will denote the $j$th character in $h$ by $h_j$, we write $\,h[i] \,$ for $\,\#M+\#I \,$ in the prefix
$\,h_1 h_2 \ldots h_i$, and we write
$\,h \langle j \rangle\,$ for $\,\#M+\#D\,$ in
the prefix $\,h_1 h_2 \ldots h_j$.
The cardinality of the set ${\cal A}_{n.m}$ of all alignments can be computed
as the coefficient of $x^m y^n$ in the generating function
$1/(1-x-y-xy)$. These coefficients are known as
 \emph{Delannoy numbers} in combinatorics
\cite[\S 6.3]{Stanley:99}. 

\begin{figure}
  \begin{center}
   \includegraphics[scale=0.7]{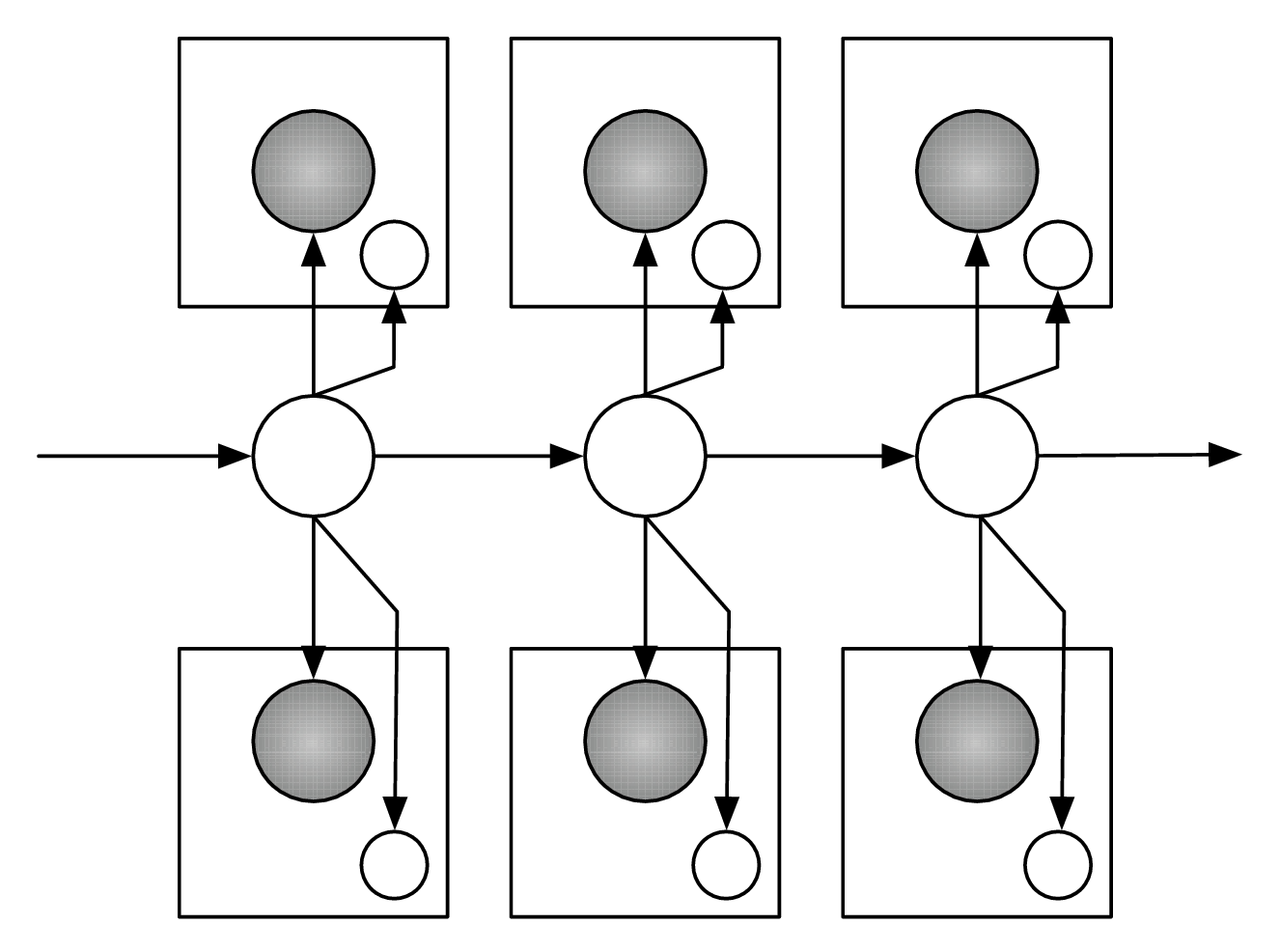}
  \end{center}
  \caption{A pair hidden Markov model for sequence alignment.}
  \label{fig:pairHMM}
\end{figure}

{\em Bayesian multi-nets} were introduced in \cite{Friedman:97} and are 
extensions of graphical models via the introduction of class nodes, and a 
set of local networks corresponding to values of the class nodes. 
In other words, the value of a random variable can change the structure 
of the graph underlying the graphical model. The 
{\em pair hidden Markov model} (see Figure \ref{fig:pairHMM}) is
 an instance of a Bayesian multinet. In this model,
the hidden states (unshaded nodes forming the chain) take on 
one of the values $M,I,D$. Depending on the value at a hidden node, 
either one or two characters are generated; this is encoded by plates (squares around the observed states) and class nodes (unshaded nodes in the plates). 
The class nodes take on the values $0$ or $1$ corresponding to whether 
or not a character is generated.
Pair hidden Markov models are 
therefore probabilistic models of alignments, in which the structure of 
the model depends on the assignments to the hidden states. 

Our next result gives the precise description of the pair HMM for sequence alignment in 
the language of algebraic statistics, namely, we represent this model
by means of a polynomial map $f$.
Let $\sigma^1$, $\sigma^2$ be the output strings from a pair hidden Markov model (of lengths $n,m$ respectively). Then:
\begin{equation}
\label{pairhmm}
f_{\sigma^1,\sigma^2}
 \quad = \quad \sum_{h \in {\cal A}_{n,m}} 
 t_{h_1}(\sigma^1_{h[1]},\sigma^2_{h \langle 1 \rangle}) \cdot
\prod_{i = 2}^{|h|}
s_{h_{i-1}h_i} \cdot t_{h_i}(\sigma^1_{h[i]},\sigma^2_{h \langle i \rangle}) ,
\end{equation}
where $s_{h_{i-1}h_i}$ is the transition probability from state $h_{i-1}$ to $h_i$ and $t_{h_i}(\sigma^1_{h[i]},\sigma^2_{h \langle i \rangle})$ are the output probabilities 
for a given state $h_i$ and the corresponding output characters on the strings $\sigma^1,\sigma^2$.

\begin{prop} \label{pairHMMmap}
The pair hidden Markov model for sequence alignment is the 
image of a polynomial map $f : {\bf R}^{9 + 2l+ l^2 } 
\rightarrow {\bf R}^{l^{n+m}}$.
The coordinates of $f$ are 
polynomials
of degree $n  +  m + 1 $ in  (\ref{pairhmm}).
\end{prop}

We need to explain why the number of parameters is $9 + 2l+ l^2 $.
First, there are nine parameters
$$ S \quad = \quad
\begin{pmatrix}
s_{MM} &  s_{MI} &  s_{MD} \\
s_{IM} &  s_{II} &  s_{ID} \\
s_{DM} &  s_{DI} &  s_{DD} 
\end{pmatrix} , $$
which play the same role as in Section 2,
namely, they represent transition probabilities
in the Markov chain. There are
$l^2$ parameters $\,t_M(a,b) =: t_{Mab}\,$
for the probability that letter $a$ in
$\sigma^1$ is matched with letter $b$ in $\sigma^2$.
The insertion parameters $\,t_I(a,b) \,$
depend only on the letter $b$, and the
deletion parameters  $\,t_D(a,b) \,$
depend only on the letter $a$, so there
are only  $2l $ of these parameters. In the upcoming example,
which explains the algebraic representation of 
Proposition \ref{pairHMMmap},
we use the abbreviations $\,t_{Ib}\,$ and $\,t_{Da}\,$
for these parameters.

Consider two sequences $\, \sigma^1 = ij \,$ and $\sigma^2 = klm \,$ 
of length $n = 2$ and $m = 3$ over any alphabet.
The number of alignments is $\,\#( {\cal A}_{n,m} ) = 25$, and they are listed in Table 1.
\begin{table}
\begin{center}
\begin{tabular} {|l|l|l|}  \hline
IIIDD & \,\, $( \,\cdot \cdot \cdot ij \,,\, klm\cdot \cdot  \, )$ & $  
 t_{Ik} s_{II} t_{Il} s_{II} t_{Im} s_{ID} t_{Di} s_{DD} t_{Dj} $\\ 
IIDID & \,\, $( \,\cdot \cdot i\cdot j \, ,\, kl\cdot m\cdot  \, )$ & $  
 t_{Ik} s_{II} t_{Il} s_{ID} t_{Di} s_{DI} t_{Im} s_{ID} t_{Dj} $\\ 
IIDDI & \,\, $( \,\cdot \cdot ij \,\cdot \,,\, kl\cdot \cdot m \, )$ & $  
 t_{Ik} s_{II} t_{Il} s_{ID} t_{Di} s_{DD} t_{Dj} s_{DI} t_{Im} $\\ 
IDIID & \,\, $( \,\cdot \, i\cdot \cdot j\,,\, k\cdot lm\cdot  \, )$ & $  
 t_{Ik} s_{ID} t_{Di} s_{DI} t_{Il} s_{II} t_{Im} s_{ID} t_{Dj} $\\ 
IDIDI & \,\, $( \,\cdot \, i\cdot j\cdot \,,\, k\cdot l\cdot m \, )$ & $  
 t_{Ik} s_{ID} t_{Di} s_{DI} t_{Il} s_{ID} t_{Dj} s_{DI} t_{Im} $\\ 
IDDII & \,\, $( \,\cdot \,ij \cdot \cdot \,,\, k\cdot \cdot lm \, )$ & $  
 t_{Ik} s_{ID} t_{Di} s_{DD} t_{Dj} s_{DI} t_{Il} s_{II} t_{Im} $\\ 
DIIID & \,\, $( \,i\cdot \cdot \cdot j \,,\, \cdot \, klm\cdot  \, )$ & $  
 t_{Di} s_{DI} t_{Ik} s_{II} 
t_{Il} s_{II} t_{Im} s_{ID} t_{Dj} $\\ 
DIIDI & \,\, $( \,i\cdot \cdot j\cdot \,,\, \cdot \,kl\cdot m \, )$ & $  
 t_{Di} s_{DI} t_{Ik} s_{II} t_{Il} s_{ID} t_{Dj} s_{DI} t_{Im} $\\ 
DIDII & \,\, $( \,i\cdot j\cdot \cdot \,,\, \cdot \,k\cdot lm \, )$ & $  
 t_{Di} s_{DI} t_{Ik} s_{ID} t_{Dj} s_{DI} t_{Il} s_{II} t_{Im} $\\ 
DDIII & \,\, $( \,ij\cdot \cdot \,\cdot\, ,\, \cdot \cdot klm \, )$ & $  
 t_{Di} s_{DD} t_{Dj} s_{DI} t_{Ik} s_{II} t_{Il} s_{II} t_{Im} $\\ 
MIID & \,\, $( \,i\cdot \cdot j \,,\, klm \,\cdot  \, )$ & $     t_{Mik} s_{MI} t_{Il} s_{II} t_{Im} s_{ID} t_{Dj} $\\ 
MIDI & \,\, $( \,i\cdot j\cdot \,,\, kl\cdot m \, )$ & $     t_{Mik} s_{MI} t_{Il} s_{ID} t_{Dj} s_{DI} t_{Im} $\\ 
MDII & \,\, $( \,ij\cdot \cdot \,,\, k\cdot lm \, )$ & $     t_{Mik} s_{MD} t_{Dj} s_{DI} t_{Il} s_{II} t_{Im} $\\ 
IMID & \,\, $( \,\cdot \,i\cdot j \,,\, klm\cdot  \, )$ & $     t_{Ik} s_{IM} t_{Mil} s_{MI} t_{Im} s_{ID} t_{Dj} $\\ 
IMDI & \,\, $( \,\cdot \,ij\,\cdot \,,\, kl\cdot m \, )$ & $     t_{Ik} s_{IM} t_{Mil} s_{MD} t_{Dj} s_{DI} t_{Im} $\\ 
IIMD & \,\, $( \,\cdot \cdot ij\,,\, klm \,\cdot  \, )$ & $     t_{Ik} s_{II} t_{Il} s_{IM} t_{Mim} s_{MD} t_{Dj} $\\ 
IIDM & \,\, $( \,\cdot \cdot ij\,,\, kl\cdot m \, )$ & $     t_{Ik} s_{II} t_{Il} s_{ID} t_{Di} s_{DM} t_{Mjm} $\\ 
IDMI & \,\, $( \,\cdot ij\cdot \,,\, k\cdot lm \, )$ & $     t_{Ik} s_{ID} t_{Di} s_{DM} t_{Mjl} s_{MI} t_{Im} $\\ 
IDIM & \,\, $( \,\cdot i\cdot j\,,\, k\cdot lm \, )$ & $     t_{Ik} s_{ID} t_{Di} s_{DI} t_{Il} s_{IM} t_{Mjm} $\\ 
DMII & \,\, $( \,ij\cdot \cdot \,,\, \cdot \,klm \, )$ & $     t_{Di} s_{DM} t_{Mjk} s_{MI} t_{Il} s_{II} t_{Im} $\\ 
DIMI & \,\, $( \,i\cdot j\cdot \,,\, \cdot \,klm \, )$ & $     t_{Di} s_{DI} t_{Ik} s_{IM} t_{Mjl} s_{MI} t_{Im} $\\ 
DIIM & \,\, $( \,i\cdot \cdot j\,,\, \cdot \,klm \, )$ & $     t_{Di} s_{DI} t_{Ik} s_{II} t_{Il} s_{IM} t_{Mjm} $\\ 
MMI & \,\, $( \,ij \,\cdot\,\, , \,\,klm \, )$ & $     t_{Mik} s_{MM} t_{Mjl} s_{MI} t_{Im} $\\ 
MIM & \,\, $( \,i \cdot j \,\,,\,\, klm \, )$ & $     t_{Mik} s_{MI} t_{Il} s_{IM} t_{Mjm} $\\ 
IMM & \,\, $( \,\cdot \,ij \,\,,\,\, klm \, )$ & $     t_{Ik} s_{IM} t_{Mil} s_{MM} t_{Mjm} $\\ \hline
\end{tabular}
\end{center}
\caption{Alignments for a pair of sequences of length $2$ and $3$.} 
\end{table}
The polynomial $f_{\sigma^1,\sigma^2}$ is the sum of the
$25$ monomials (of degree $9,7,5$) in the rightmost column.
For instance, if we consider strings over the binary
alphabet $\{0,1\}$, then there are $17$ parameters
(nine $s$-parameters and eight $t$-parameters), and
the pair HMM for alignment is the image of a map
$ \, f : {\bf R}^{17} \rightarrow {\bf R}^{32}$.
The coordinate of $f$ which is indexed by
$(i,j,k,l,m)  \in \{0,1\}^5$ equals the
$25$-term polynomial gotten by summing the 
rightmost column in Table 1.

The parametric inference problem for sequence alignment is solved 
by computing the Newton polytopes $NP(f_{\sigma_1,\sigma_2})$ with the
polytope propagation algorithm. 
In the terminology introduced in \cite[\S 4]{Pachter:04},
an observation $\sigma$ in the pair HMM is the pair of sequences
$(\sigma_1,\sigma_2)$, and the possible explanations
are the optimal alignments of these sequences with
respect to the various choices of parameters.
In summary, the vertices of the Newton polytope
$NP(f_{\sigma_1,\sigma_2})$ correspond to the optimal alignments.
If the observed sequences $\sigma_1,\sigma_2$ are not fixed then we are in the situation of
\cite[Proposition 6]{Pachter:04}.
 Each parameter choice
defines a function from pairs of sequences to alignments:
$$\, \{0,\ldots,l-1\}^n \times \{0,\ldots,l-1\}^m
\rightarrow {\cal A}_{n,m} \,,\quad ( \sigma_1,\sigma_2) \mapsto \hat {\bf h}  .$$
The number of such functions
grows doubly-exponentially in $n$ and $m$, but only 
a tiny fraction of them are \emph{inference functions},
which means they correspond to the vertices of the Newton polytope
of the map $f$.
It is an interesting combinatorial problem to characterize
the inference functions for sequence alignment.

An important observation is that our formulation in Problem 2 is equivalent to 
combinatorial ``scoring schemes'' or ``generalized edit distances'' which 
can be used to assign weights to alignments \cite{Bucher:96}. 
For example, the simplest scoring scheme consists of two parameters:
 a mismatch score $mis$, and an indel score $gap$ \cite{Fernandez-Baca:00, Gusfield:94, Waterman:92}. 
The weight of an alignment is the sum of the scores for all positions in the alignment, where a match is assigned a score of $1$. 
This is equivalent to specializing the logarithmic parameters
$U = - {\rm log} (S)$ and $V = - {\rm log} (T)$ of the pair hidden Markov model as follows:
\begin{equation}
\label{specialize}
u_{ij} = 0, \quad
v_{Mij}=1 \,\hbox{ if $i=j$}, \,\,\,\,
v_{Mij}=mis\, \hbox{ if $i \neq j$, and }\,\,\,\,
v_{Ij} =  v_{Di} = gap
\qquad \hbox{for all $i,j$}. 
\end{equation}
This specialization of the parameters
corresponds to intersecting the normal fan of
the Newton polytope with a two-dimensional affine subspace
(whose coordinates are called $mis$ and $gap$).

Efficient software for parametrically aligning the sequences with two free parameters
already exists (XPARAL \cite{Gusfield:96}).
Consider the example of the following two sequences:
$\sigma^1=AGGACCGATTACAGTTCAA$ and $\sigma^2=TTCCTAGGTTAAACCTCATGCA$. XPARAL will return four cones, however a computation of the Newton polytope reveals seven vertices (three correspond to positive $mis$ or $gap$ values). The polytope propagation algorithm has 
the same running time as XPARAL: for two sequences of 
length $n,m$, the method requires $O(nm)$ two-dimensional convex hull computations. The number of points in each computation is bounded by the total
number of points in the final convex hull (or equivalently the number, $K$, of explanations). Each convex hull computation therefore
requires at most $O(K {\rm log}(K))$ operations, thus giving an $O(nmK {\rm log}(K))$ algorithm for solving the parametric alignment problem. However, this 
running time can be improved by observing that the convex hull computations that need to be carried out have a very special form, namely in each 
step of the algorithm we need to compute the convex hull of two superimposed convex polygons. This procedure is in fact a primitive of the divide
and conquer approach to convex hull computation, and there is a well known $O(K)$ algorithm for solving it  \cite[\S 3.3.5]{Preparata:85}. Therefore, for two parameters, our recursive approach
to solving the parametric problem yields an $O(Kmn)$ algorithm, matching the running time of XPARAL and the conjecture of Waterman, Eggert and Lander \cite{Waterman:92}.

\begin{figure}[ht]
  \begin{center}
   \includegraphics[scale=1.2]{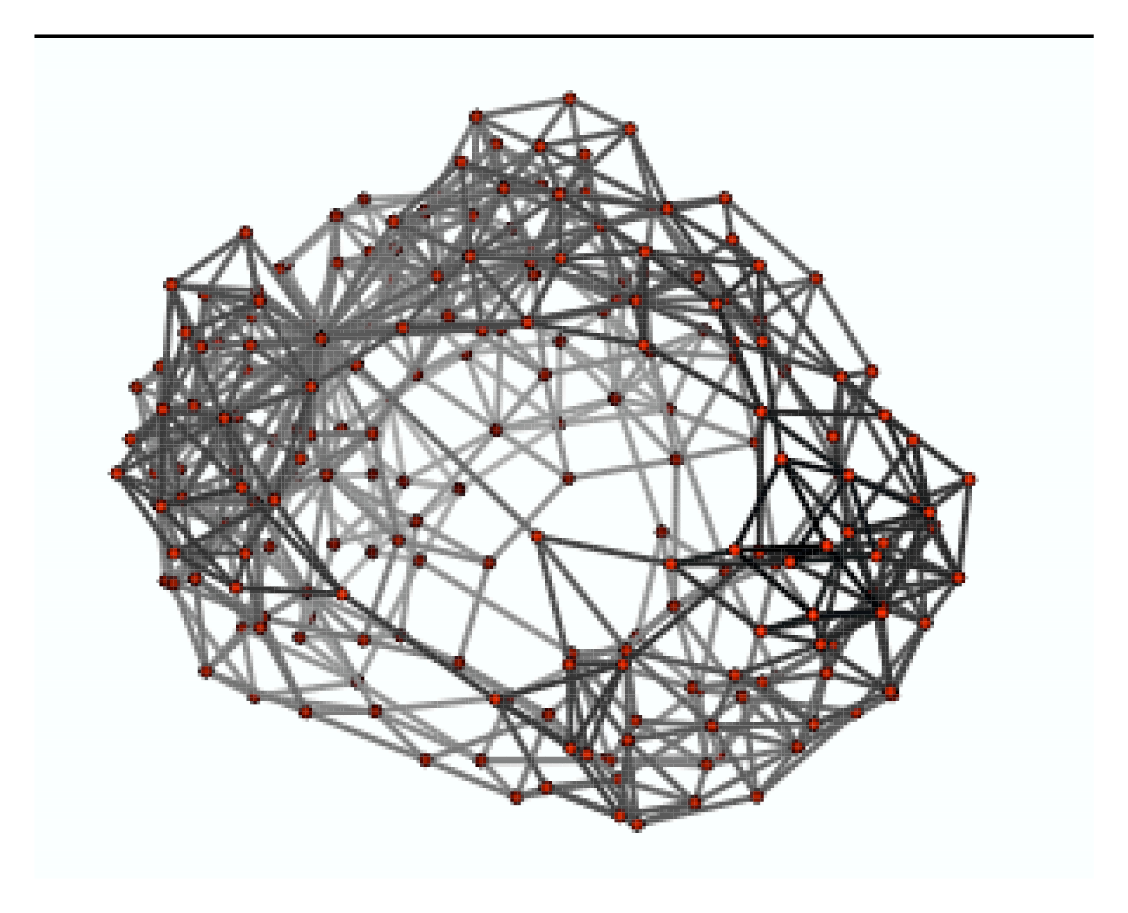}
  \end{center}
  \caption{Edge graph of the Newton polytope for a four parameter alignment problem.}
  \label{fig:parametric}
\end{figure}

In order to demonstrate the practicality of our approach for higher-dimensional problems, we implemented a four parameter recursive parametric alignment solver. The more 
general alignment model includes different transition/transversion parameters (instead of just one mismatch parameter), and separate parameters for 
opening gaps and extending gaps. A transition is mutation from one purine ($A$ or $G$) to another, or from one pyrimidine ($C$ or $T$) to another, and a transversion is a mutation
from a purine to a pyrimidine or vice versa. More precisely, if we let $P_u=\{A,G\}$ and $P_y=\{C,T\}$ the model is:
\begin{eqnarray*}
\label{specialize2}
u_{MM} = u_{IM} = u_{DM} & = &  0\\
u_{MI} = u_{MD} & = & gapopen\\
u_{II} = u_{DD} & = & gapextend\\
v_{Mij} & = & \hbox {$1$ if $i=j$}\\
v_{Mij} & = & transt\, \  \hbox {if $i \neq j$, and $i,j \in P_u$ or $i,j \in P_y$}\\
v_{Mij} & = & transv\, \ \hbox {if $i \neq j$, and $i \in P_u, j \in P_y$ or vice versa}\\
v_{Ij} =  v_{Di} & = & \hbox{$0$ for all $i,j$}.
\end{eqnarray*}

For the two sequences $\sigma^1$ and $\sigma^2$ in the example above, the number of vertices of the four dimensional 
Newton polytope (shown in Figure 4) is $224$ (to be compared to $7$ for the two parameter case).

\section{Practical Aspects of Parametric Inference}

We begin by pointing out that parametric inference is useful for Bayesian computations. Consider the problem where we have a prior distribution $\pi(s)$ on our parameters
$s = (s_1,\ldots,s_d)$, and we would like to compute the posterior probability of a maximum a posteriori explanation $\widehat {\bf h}$:
\begin{equation}
\label{Bayesian}
{\rm Prob}({\bf X} = \widehat {\bf h} \,|\, {\bf Y} = {\bf \sigma}) \quad 
= \quad \int_{s} {\rm Prob}({\bf X}= \widehat {\bf h} \,|\, {\bf Y}
 = {\bf \sigma},\,s_1,\ldots,s_d \,)\pi(s)  ds. 
\end{equation}
This is an important problem, since it can give a quantitative assessment of the validity of $\widehat {\bf h}$ in a setting where we have prior, but not certain, information about the parameters, and also because we may want to sample $\widehat {\bf h}$ according to its posterior distribution (for an example of how this can be applied in computational biology see \cite{Liu:94}). Unfortunately, these integrals may be difficult to compute. We propose the following simple 
 Monte Carlo algorithm for computing a numerical approximation
 to the integral (\ref{Bayesian}):

\begin{prop}
Select $N$ parameter vectors $s^{(1)},\ldots,s^{(N)}$
 according to the distribution $\pi(s)$, where
 $N$ is much larger than the
 number of vertices of the Newton polytope $\,NP(f_\sigma)$.
 Let $K$ be the number of $s^{(i)}$ 
 such that $-{\rm log}(s^{(i)})$ lies in the normal cone of
$NP(f_\sigma)$ indexed by the explanation $\widehat {\bf h}$.
 Then $K/N$ approximates (\ref{Bayesian}).
\end{prop}

 \begin{proof}
 The expression $\,
  {\rm Prob}({\bf X}= \widehat {\bf h} \,|\, {\bf Y}
 = {\bf \sigma},\,s_1,\ldots,s_d \,) \,$ is
 zero or one depending on whether the vector
 $-{\rm log}(s) = (-{\rm log}(s_1),\ldots,-{\rm log}(s_d))
 $ lies in the normal cone of
$NP(f_\sigma)$ indexed by $\widehat {\bf h}$. This membership test can be done without ever running the sum-product algorithm if we precompute an inequality representation of the normal cones.
\end{proof}

The bound on the number of vertices of the Newton polytope
in \cite[\S 4]{Pachter:04} provides a valuable tool for
estimating the quality of this Monte Carlo approximation.
We believe that the tropical geometry developed in \cite{Pachter:04}
will also be useful for more refined analytical approaches to
Bayesian integrals.  The study of Newton polytopes
can also complement the algebraic geometry
approach to model selection proposed in \cite{Rusakov:02}.

Another application of parametric inference is to problems where the number of parameters may be very large, but where we want to fix a large subset of them, thereby reducing the dimensions of the polytopes. Gene finding models, for example, may have up to thousands of parameters and input sequences can be millions of base pairs long however, we are usually only interested in studying the dependence of inference on a select few. Although specializing parameters reduces the dimension of the parameter space, the explanations correspond to vertices of a 
\emph{regular subdivision of the Newton polytope}, rather than just to the vertices of the polytope itself. This is explained below (readers may also 
want to refer to \cite{Pachter:04} for more background).

Consider a graphical model with parameters $s_1,\ldots, s_{d}$
 of which the parameters $s_1,\ldots , s_{r}$ are
free but $\, s_{r+1} = S_{r+1}, \ldots, s_d = S_d \,$
where the $S_i$ are fixed non-negative numbers.
Then the coordinate polynomials $f_\sigma$ of our model 
specialize to polynomials in $r$ unknowns
whose coefficients $c_a$ are non-negative numbers:
$$\, \tilde f_\sigma(s_1,\ldots,s_{r}) \quad = \quad
 f_\sigma(s_1,\ldots,s_{r}, S_{r+1}, \ldots, S_{d})
\quad = \quad \sum_{a \in {\bf N}^r} c_a \cdot s_1^{a_1} \cdots s_{r}^{a_r}. $$
The \emph{support} of this polynomial is the finite set
$\, {\cal A}_\sigma \, = \, \{\, a \in {\bf N}^r \, : \,c_a > 0 \,\}$.
The convex hull of   $\, {\cal A}_\sigma\,$ in ${\bf R}^r$
is the  Newton polytope of the polynomial $\tilde f_\sigma  = \tilde f_\sigma(s_1,\ldots,s_r)$. For example, in the case of the hidden Markov model with output parameters specialized,
the Newton polytope of
 $\tilde f_{\sigma}$ is the polytope associated with a Markov chain.
  Kuo \cite{Kuo:04} shows that the size of these
  polytopes does not depend on the length of the chain. 

Let ${\bf h}$ be any explanation for $\sigma$ in the original model
and let $(u_1,\ldots,u_r,u_{r+1}, \ldots,u_n)$ be the vertex
of the Newton polytope of $f_\sigma$ corresponding
to that explanation. We abbreviate $\,a_{\bf h} = (u_1,\ldots,u_r)\,$
and $\, S_{\bf h} \, = \,S_{r+1}^{u_{r+1}} \cdots S_d^{u_d}$.
The assignment
$\, {\bf  h} \mapsto a_{\bf h}\,$ defines  a map
from the set of explanations of $\sigma$ to the support
$\, {\cal A}_\sigma$. The convex hull of
the image coincides with the Newton polytope of $\,\tilde f_\sigma$.
We define
\begin{equation}
\label{fromHtoA}
 w_a \, = \, {\rm min} \bigl\{ \, - {\rm log}(S_{\bf h}) \, \, :\,\,
{\bf h} \, \, \hbox{is an explanation for } \, \sigma \,\,\, \hbox{with}\,\,\,
a_{\bf h} = a \, \bigr\}.  
\end{equation}
If the specialization is sufficiently generic
then this maximum is attained uniquely,
and, for simplicity, we will assume that this is the case.
If a point $a \in {\cal A}_\sigma$ is not the image of any explanation ${\bf h}$ then 
we set $w_a = \infty$.
The assignment $a \mapsto w_a$ is a real valued function
on the support of our polynomial $\tilde f_\sigma$, 
and it defines a \emph{regular polyhedral subdivision} $\, \Delta_w \,$
of the Newton polytope $NP(\tilde f_\sigma)$. Namely, $\Delta_w$ is the polyhedral
complex  consisting of all lower faces of the polytope gotten by taking the
convex hull of the points $(a,w_a)$ in ${\bf R}^{r+1}$.
  See \cite{Sturmfels:96} for details on regular triangulations
  and regular polyhedral subdivisions.

\begin{thm}
The explanations for the observation $\sigma$ in the specialized model are
in bijection with the vertices of the regular polyhedral subdivision $\, \Delta_w \,$
of the Newton polytope of the specialized polynomial $\, \tilde f_\sigma$.
\end{thm}

\begin{proof}
The point $(a,w_a)$ is a vertex of $\Delta_w$ if and only if
the following open polyhedron is non-empty:
$$  P_a \quad = \quad \bigl\{ \, v \in {\bf R}^r \,\, : \,\,
 a \cdot v + w_a \, < \, a' \cdot v + w_{a'}\,\, \hbox{for all}\,\,
a \in {\cal A}_\sigma \backslash \{a\} \, \bigr\}. $$
If $v$ is a point in  $P_a$ then we set
$\,s_i = {\rm exp}(-v_i)\,$ for $i=1,\ldots,r$,
  and we consider the explanation ${\bf h}$
which attains the minimum in  (\ref{fromHtoA}).
Now all parameters have been specialized
and ${\bf h}$ is the solution to Problem 2.
This argument is reversible: any explanation for 
$\sigma$ in the specialized model arises from 
one of the non-empty polyhedra $P_a$.
  We note that the collection of polyhedra $P_a$ defines a polyhedral 
subdivision of ${\bf R}^r$ which is geometrically dual 
to the subdivision $\Delta_w$ of the Newton polytope
of $\tilde f_\sigma$. 
\end{proof}

 \vskip .1cm

 In practical applications of parametric inference, it
 may be of interest to compute only one normal cone of the Newton polytope (for example the cone containing some fixed parameters). We conclude this section by observing that the polytope propagation algorithm is suitable for this computation as well:

\begin{prop}
Let $v$ be a vertex of a $d$-dimensional Newton polytope of a hidden Markov model. Then the normal cone containing $v$ can be computed using a polytope propagation algorithm
in dimension $d-1$.
\end{prop}

\begin{proof} We run the standard polytope propagation algorithm
described in Section 4,
but at each step we record only the minimizing vertex in the direction of the log parameters, together with its neighboring vertices in the edge graph of the Newton polytope. It follows, by induction, that given this information at the $n$th step, we can use it to find the minimizing vertices and related neighbors in the $(n+1)$st step.
\end{proof}

\section{Summary}

We envision a number of biological applications for the polytope propagation algorithm, including:

\begin{itemize}
\item Full parametric inference using the normal fan of the Newton polytope of an observation when the graphical model under 
consideration has only few model parameters. 
\item Utilization of the edge graph of the polytope  to identify stable parts of 
alignments and annotations. 
\item Construction
of the normal cone containing a specific parameter vector
when computation of the full Newton polytope is infeasible.
\item Computation of the posterior probability 
 (in the sense of Bayesian statistics) of an alignment
or annotation. The regions for the relevant integrations
are the normal cones of the Newton polytope.
\end{itemize}

As we have seen, the computation of Newton polytopes for (interesting) graphical models is certainly feasible for a few free parameters, and we expect that further analysis of the computational geometry should yield efficient algorithms in higher dimensions. For example, the key operation, computation of convex hulls of unions of convex polytopes, is likely to be considerably easier than general convex hull computations even in high
dimensions. Fukuda, Liebling and L\"{u}tlof \cite{Fukuda:01} give a polynomial time algorithm for computing extended convex hulls (convex hulls of unions of convex polytopes) under 
the assumption that the polytopes are in general position. Furthermore, it should be possible to optimize the geometric algorithms for specific models of interest, and combinatorial analysis of the Newton polytopes arising in graphical models should yield better complexity estimates (see, e.g., \cite{ Fernandez-Baca:00, Gusfield:94}).
 Michael Joswig is currently working on a general polytope propagation implementation in POLYMAKE \cite{Gawrilow:00,Gawrilow:01}.

In the case where computation of the Newton polytope is impractical, it is still possible to identify the cone containing a specific parameter, and this can be used to quantitatively measure the robustness of the inference. Parameters near a boundary are unlikely to lead to biologically meaningful results. Furthermore, the edge graph can be used to identify common regions in the explanations corresponding to adjacent vertices. In the case of alignment, biologists might see a collection of alignments rather than just one optimal one, with common sub-alignments highlighted. This is quite different from returning the $k$ best alignments, since suboptimal alignments may not be vertices of the Newton polytope. The solution we propose explicitly identifies all suboptimal alignments that can result from similar parameter choices. 

\section{Acknowledgments} 
Lior Pachter was supported in part by a grant from the NIH (R01-HG02362-02).
Bernd Sturmfels was supported by
a Hewlett Packard Visiting Research Professorship 2003/2004
at MSRI Berkeley and in part  by the NSF (DMS-0200729).


\nocite{*}
\begin{thebibliography}{26}

\bibitem{Alexandersson:03} M. Alexandersson, S. Cawley and L. Pachter: SLAM - Cross-species Gene Finding and Alignment with a Generalized Pair Hidden Markov Model, Genome Research 13 (2003) 496--502.
\bibitem{Baldi:98} P.  Baldi and S.  Brunak:   Bioinformatics. The Machine Learning Approach. A Bradford Book.The MIT  Press. Cambridge, Massachusetts, 1998.
\bibitem{Bucher:96} P. Bucher and K. Hofmann: A sequence similarity 
search algorithm based on a probabilistic interpretation of an alignment 
scoring system, Proceedings of the Conference on Intelligent Systems for Molecular Biology, 1996,  44--51.
\bibitem{Durbin:98} R. Durbin, S. Eddy, A. Krogh and G. Mitchison: Biological Sequence Analysis (Probabilistic Models of Proteins and Nucleic Acids), 
Cambridge University Press, 1998.
\bibitem{Fernandez-Baca:00} D. Fern\'andez-Baca, T. Sepp\"al\"ainen and G. Slutzki:
Parametric multiple sequence alignment and phylogeny construction, 
in Combinatorial Pattern Matching, Lecture 
Notes in Computer Science (R. Giancarlo and D. Sankoff eds.), Vol. 1848, 2000, 68--82.
\bibitem{Friedman:97} N. Friedman, D. Geiger and M. Goldszmidt: Bayesian network classifiers, Machine Learning 29 (1997) 131--161.
\bibitem{Fukuda:01} K. Fukuda, T.H. Liebling and C. L\"{u}tlof: Extended convex hull, Computational Geometry  20 (2001) 13--23.
\bibitem{Gardiner-Garden:87} M. Gardiner-Garden and M. Frommer: CpG islands in vertebrate genomes, Journal of Molecular Biology 196 (1987) 261--282.
\bibitem{Gawrilow:00} E. Gawrilow and M. Joswig: polymake: a Framework for Analyzing Convex Polytopes, Polytopes -- Combinatorics and Computation (G. Kalai and G.M. Ziegler eds.), Birkhh\"{a}user (2000).
\bibitem{Gawrilow:01} E. Gawrilow and M. Joswig: polymake: an Approach to Modular Software Design in Computational Geometry, Proceedings of the 17th Annual Symposium on Computational Geometry, ACM, 2001, 222--231.
\bibitem{Gusfield:94} D. Gusfield, K. Balasubramanian, and D. Naor: Parametric optimization of sequence alignment, Algorithmica 12 (1994) 312--326.
\bibitem{Gusfield:96} D. Gusfield and P. Stelling: Parametric and inverse-parametric sequence alignment with XPARAL, Methods Enzymology 266  (1996) 481--494.
\bibitem{Jordan:02} M.I. Jordan and Y. Weiss: Graphical Models: 
Probabilistic Inference, in {\it Handbook of Brain Theory and 
Neural Networks, 2nd edition}, M. Arbib (Ed.), Cambridge, MA, MIT 
Press, 2002.
\bibitem{Kschischang:01} F. Kschischang, B. Frey, and H. A. Loeliger: Factor graphs and the sum-product algorithm, IEEE Trans. Inform. Theory 47 (2001) 498--519.
\bibitem{Kuo:04} E. Kuo, Viterbi sequences and polytopes, {\tt http://front.math.ucdavis.edu/math.CO/0401342}.
\bibitem{Lander:01} E.S. Lander et al.: Initial sequencing and analysis of the human genome, Nature 409 (2001) 860--921.
\bibitem{Liu:94} J. Liu: The collapsed Gibbs sampler with applications to a gene regulation problem, J. Amer. Statist. Assoc.~89 (1994) 958--966. 
\bibitem{Pachter:04} L. Pachter and B. Sturmfels: Tropical geometry of statistical models,  companion paper, submitted.
\bibitem{Preparata:85} F. P. Preparata and M. I. Shamos: Computational Geometry- An Introduction, Springer Verlag 1985.
\bibitem{Rusakov:02}  D.~Rusakov and D.~Geiger: Asymptotic model
selection for naive Bayesian networks, Uncertainty in
Artificial Intelligence, 2002, 438--445. 
\bibitem{Stanley:99} R. Stanley: Enumerative Combinatorics, Volume 2,
Cambridge University Press, 1999.
\bibitem{Sturmfels:96} B. Sturmfels: Gr\"obner Bases and Convex Polytopes, University Lecture Series, Vol. 8, American Mathematical Society, 1996.
\bibitem{Takai:02} D. Takai and P. A. Jones: Comprehensive analysis of CpG islands in human chromosomes 21 and 22, Proc. Natl. Acad. Sci. USA 99 (2002) 3740--3745.
\bibitem{Waterman:92} M. Waterman, M. Eggert and E. Lander: Parametric 
sequence   comparisons, Proc. Natl. Acad. Sci. USA 89 (1992) 6090--6093.
\end{thebibliography}
\end{document}